\documentclass{article}
\usepackage[legalpaper, margin=1.5in]{geometry}
\usepackage[utf8]{inputenc}
\usepackage{amsfonts,amsmath,amssymb,amsthm,amstext,latexsym,stmaryrd}

\usepackage{thmtools}
\usepackage{thm-restate}
\usepackage{hyperref}
\declaretheorem{theorem}
\usepackage{url}
\usepackage{xspace}
\usepackage[nocompress]{cite}
\usepackage{array}
\usepackage{ifthen}
\usepackage{intcalc}
\usepackage{algpseudocode}
\usepackage{subcaption}
\usepackage[pdftex,dvipsnames]{xcolor}
\usepackage{mathtools}
\usepackage{multirow,arydshln}
\usepackage[normalem]{ulem}
\usepackage{cleveref}

\newtheorem{definition}[theorem]{Definition}

\newtheorem*{note*}{Note}

\newtheorem*{remark*}{Remark}

\newtheorem*{claim*}{Claim}

\definecolor{darkgreen}{rgb}{0,0.5,0}
\definecolor{midnight}{rgb}{0,0.094,0.533}
\definecolor{ocean}{rgb}{0,0.290,0.533}

\usepackage[algo2e,ruled, vlined, linesnumbered]{algorithm2e}
\SetKwInput{KwInput}{Input}
\SetKw{Continue}{continue}

\newcommand{\Pred}[1]{{\tt Pred}[#1]\xspace}
\newcommand{\Succ}[1]{{\tt Succ}[#1]\xspace}

\newcommand{\Neigh}[1]{{\tt Neigh}[#1]\xspace}

\title{A Simple and Elegant Mathematical Formulation for the Acyclic DAG Partitioning Problem}
\date{}

\author{M. Yusuf \"Ozkaya\\
Georgia Institute of Technology\\
School of Computational Science and Engineering\\
\url{https://myozka.github.io}\\
\and
\"Umit V. \c{C}ataly\"urek\footnote{This publication describes work performed at the Georgia
Institute of Technology and is not associated with Amazon.}\\
Georgia Institute of Technology\\
School of Computational Science and Engineering\\
and\\
Amazon Web Services\\
\url{https://faculty.cc.gatech.edu/~umit/}}

\begin{document}

\maketitle

\begin{abstract}
This work addresses the NP-Hard problem of acyclic directed acyclic graph (DAG) partitioning problem.
The acyclic partitioning problem is defined as partitioning the
vertex set of a given directed acyclic graph into disjoint and collectively exhaustive subsets (parts).
Parts are to be assigned such that the total sum of the vertex weights within each part satisfies a
common upper bound and the total sum of the edge costs that connect nodes across different parts is
minimized. Additionally, the quotient graph, i.e., the induced graph where all nodes that are assigned
to the same part are contracted to a single node and edges of those are replaced with cumulative
edges towards other nodes, is also a directed acyclic graph. That is, the quotient graph itself is
also a graph that contains no cycles. Many computational and real-life applications such as
in computational task scheduling, RTL simulations, scheduling of rail-rail transshipment tasks and
Very Large Scale Integration (VLSI) design make use of acyclic DAG partitioning.
We address the need for a simple and elegant mathematical formulation for the
acyclic DAG partitioning problem that enables easier understanding, communication, implementation,
and experimentation on the problem.
\end{abstract}

\section{Introduction}
\label{sec.intro}

Graph partitioning has been an active area of research for several decades, and is an essential
technique for
data and computation distribution for efficient computation~\cite{kaku:98:metis,Catalyurek99,HENDRICKSON20001519}.
A graph partitioning problem is, in general, defined as the task of dividing the vertex set of a
directed or undirected graph into roughly balanced disjoint subsets (also called as parts,
partitions, and clusters)
while minimizing the total weight of edges that connect nodes from different parts~\cite{pell:08:scotch,kaku:98:metis,nossack2014mathematical}.
Research community has developed various types of graphs for the many specific real-world scenario or applications.
One type of graphs that is perhaps the de-facto for task scheduling and task/workflow management is
directed acyclic graphs (DAGs). A directed acyclic graph can be defined as a graph in which there
are no cycles, i.e., if there is a directed path between from a node $A$ to a node $B$,
there is no path from node $B$ to node $A$. Hence, many efforts have focused on partitioning
this specific subclass of graphs.

In this work, we focus on one special constraint variant of DAG partitioning:
The acyclic DAG partitioning problem.
The acyclic DAG partitioning problem introduces an additional constraint to a regular graph partitioning
problem: The quotient graph of the partitioning, i.e., the resulting graph when all nodes that are
assigned to the same part are contracted together to a single vertex and in which the edges represent the
cumulative edges between those nodes, is also a directed acyclic graph~
\cite{mops:17b,Herrmann19-SISC,nossack2014mathematical,albareda2019reformulated}.

The acyclic DAG partitioning problem is defined and attempted to be tackled in many domains for many years
such as partitioning and computation of boolean networks~\cite{colb:94}, VLSI design~
\cite{kocan2005}, rail-rail transshipment~\cite{nope:14,albareda2019reformulated},
RTL simulations~\cite{beamer2020efficiently}, task scheduling problems~
\cite{Ozkaya19-PPAM,Ozkaya19-IPDPS}, quantum circuit simulation~\cite{Fang22-ARXIV}, etc.
Similar to graph partitioning, acyclic DAG partitioning is an NP-Hard problem~\cite{Herrmann19-SISC}.
Thus, many solutions are heuristic algorithms~\cite{mops:17b,Herrmann19-SISC}.
Most of the recent work on acyclic partitioning has either used very restrictive approaches to avoid
cycles, or expensive algorithms to detect and eliminate possible emergence of cycles~\cite{mops:17b,Herrmann19-SISC}.
Although many researchers tend towards fast heuristics for computations on larger data,
there are efforts from many domains to formulate and solve this problem to optimality using mixed
integer linear programming (ILP) models~\cite{nope:14,nossack2014mathematical,albareda2019reformulated}.

Previous work on the mathematical formulation of acyclic partitioning problem uses subtour
elimination constraints derived from Traveling Salesman Problem (TSP)~\cite{miller1960integer} or
complex formulations with possibly expensive pre-processing phases,
which may be hard to follow and discouraging for the less-informed on the topic.
Furthermore, simpler model formulations tend to be easier to approach and implement and thus,
help bridge the gap between theory, the understanding and communication of it,
and the practice in different domains.
In this work, we present a simple and elegant formulation for
optimally solving the acyclic graph partitioning problem.
Our main focus is elegantly modelling the acyclicity constraint since it is the
major source of the complexity in the formulation.
Even so, we present a concise but complete formulation for the problem.

In~\cref{sec:preliminaries}, we define the preliminaries and formally introduce the graph
partitioning and acyclic DAG partitioning problems. In~\cref{sec.related}, we describe the previous
mathematical formulations for the acyclic DAG partitioning problem and in~\cref{sec.algo}, we
present our concise and straightforward model.
In~\cref{sec:real_world_impact}, we give two example application areas where the proposed ILP
formulation can be useful.
And, in~\cref{sec.conc} we summarize and conclude the article.

\section{Preliminaries} 
\label{sec:preliminaries}

A {\em simple directed graph} $G^s = (V,E)$  contains a set of vertices $V$ and a set of directed
edges $E = \{e=(u,v) | u,v \in V \}$ for distinct $u,v \in V$, where $e$ is directed from $u$ to $v$.
That is, every edge $e = (u,v)$ connects a distinct pair of vertices. The set of edges is defined
as a subset of the cartesian of the vertex set by itself, $E \subset V \times V$.
In a {\em directed graph} $G$, in addition to the above,
every vertex $u$ has a weight associated with it
and denoted by $w_u$ and every edge $(u,v) \in E$ has a cost denoted by $c_{uv}$,
$G = (V,E,w,c)$.

A {\em path} is a sequence of edges $(u_1,v_1) \cdot (u_2,v_2),\ldots$ with $v_i=u_{i+1}$.
A path $((u_1,v_1) \cdot (u_2,v_2)  \cdot (u_3,v_3) \cdots (u_\ell,v_\ell))$ is of length $\ell$,
which connects a sequence of $\ell+1$ vertices
$(u_1, v_1 = u_2, \ldots, v_{\ell-1} = u_\ell, v_\ell)$.
A path is called {\em simple} if the connected vertices are distinct.

Let $u\leadsto v$ denote a simple path that starts from $u$ and ends at $v$.
A path $((u_1,v_1) \cdot (u_2,v_2)  \cdots (u_\ell,v_\ell))$ forms a (simple) {\em cycle} if all
$v_i$ for $1 \leq i \leq \ell$ are distinct and $u_1 = v_\ell$.
A {\em directed acyclic graph (DAG)}, is a directed graph with no cycles.
$\Pred{u} = \{ v\ |\  (v,u) \in E\}$ represent
the immediate predecessors of a vertex $u$, and $\Succ{u} = \{ v\ |\ (u,v) \in E\}$ represents the immediate successors of $v$.
The union of the immediate predecessors and successors of a vertex $u$ is called the
neighbor set of $u$: $\Neigh{u} = \Pred{u} \cup \Succ{u}$.
For a vertex $u$, the set of vertices $v$ such that there exist a path $u \leadsto v$ is called the {\em descendants} of $u$.
Similarly, the set of vertices $v$ such that $v \leadsto u$ exists is called the {\em ancestors} of the vertex $u$.
We define $N_u^v$ as the set of all distinct vertices that lie on any simple path between $u$ and
$v$, which is equivalent to the intersection of the descendant set of $u$ and the ancestor set of $v$.

\begin{definition}{A $k$-way partitioning.}
   \rm A $k$-way partitioning of a graph $G$ is a partition of its vertex set into $k$ parts (subsets)
$P~=~\{V_1, V_2, \dots, V_k\}$ such that all subsets $V_i$ are disjoint and collectively exhaustive,
$V~=~\bigcup\limits_{i=1}^{k} V_{i}$.
\end{definition}

An edge $e = (u,v)$ where $u \in V_i$, $v \in V_j$, and $i \neq j$ is called a \emph{cut edge}. The
sum of the cost of all cut edges is called \emph{edge cut}, which is a typical objective function to
minimize in the graph partitioning problems.

\begin{definition}{A balanced $k$-way partitioning.}
   \rm A $k$-way partitioning of a graph $G$ in which the total weight of vertices for a given
   subset $V_i$,
   $w(V_i) = \sum_{u \in V} w(u)$, is bounded by a small imbalance parameter $\epsilon$ by the
   inequality:
   $c(V_i) \leq (1+\epsilon)\left\lceil \frac{C(V)}{k} \right\rceil = B$.
\end{definition}

\begin{definition}{The balanced acyclic $k$-way partitioning problem.}
  \rm A balanced acyclic k-way partition $P = \{V_1,\ldots,V_k\}$ of a directed graph $G = (V, E)$ is a
  balanced $k$-way partition in which no two paths $u \leadsto v$ and $v' \leadsto u'$
  co-exist for any $\{u, u'\} \subset V_i$, $\{v, v'\} \subset V_j$, $i\neq j$. The balanced acyclic
  $k$-way partitioning problem is to find a valid balanced acyclic $k$-way partitioning of $G$ where
  the objective function (i.e., edge cut) is minimized.
\end{definition}

A $k$-way partition induces a contracted graph $\mathcal{G'} = (\mathcal{V'}, \mathcal{E'})$, i.e.,
part graph, where the nodes represent the $k$ parts of $G$, $w(u') = C(V_u), u' \in \mathcal{G'}$,
and the edges $(u', v') \in \mathcal{E'}, u' \neq v'$ represent the cumulative sum of directed edge costs
between $u'$ and $v'$, and $c_{u'v'} = \sum\limits_{u \in u', v \in v'} {c_{uv}}$.

\begin{figure}
\centering
 \begin{subfigure}[t]{.22\linewidth} \includegraphics[width=0.95\textwidth]{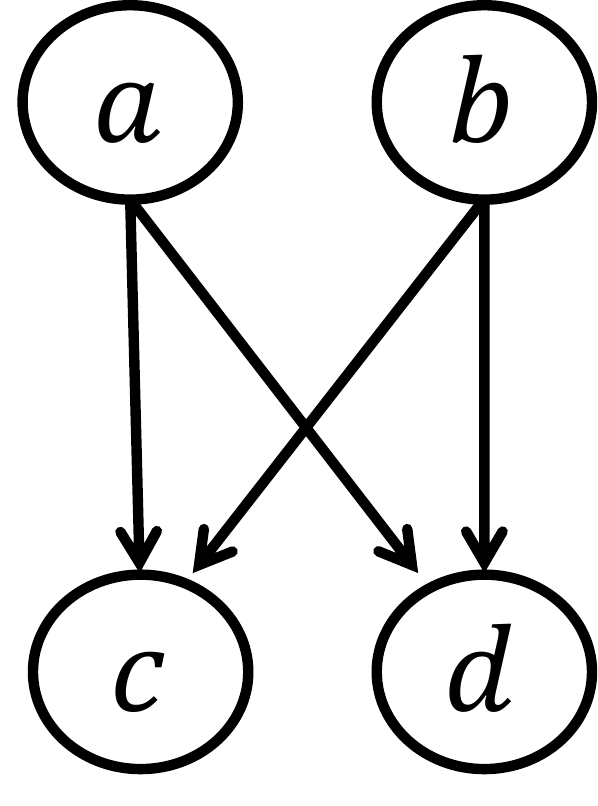}\caption{A toy graph}\label{fig:conv:1}\end{subfigure}
 \begin{subfigure}[t]{.24\linewidth} \includegraphics[width=\textwidth]{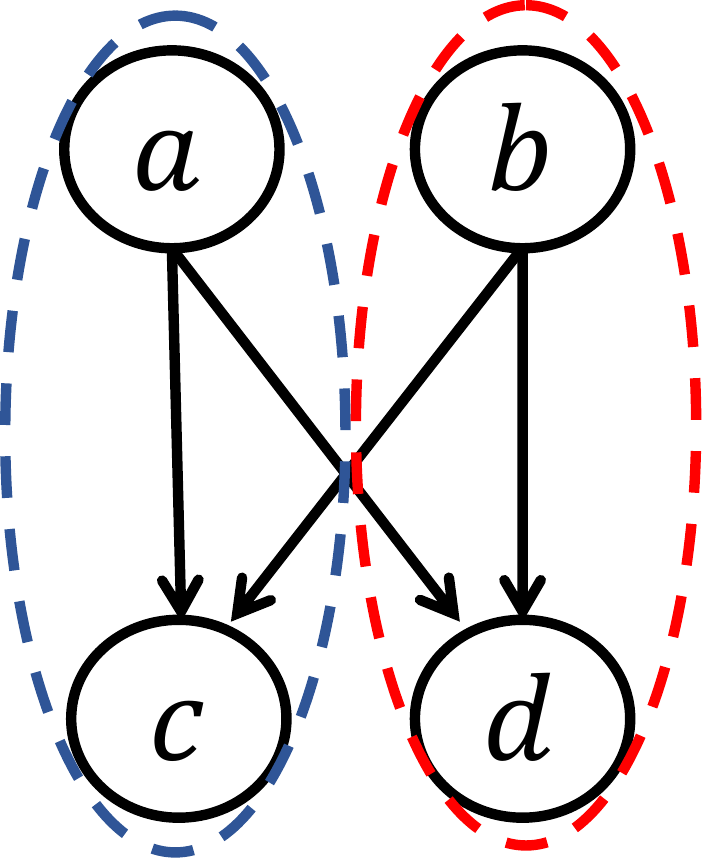}
 \caption{A cyclic partition}\label{fig:conv:2}\end{subfigure}
 \begin{subfigure}[t]{.24\linewidth} \includegraphics[width=0.9\textwidth]{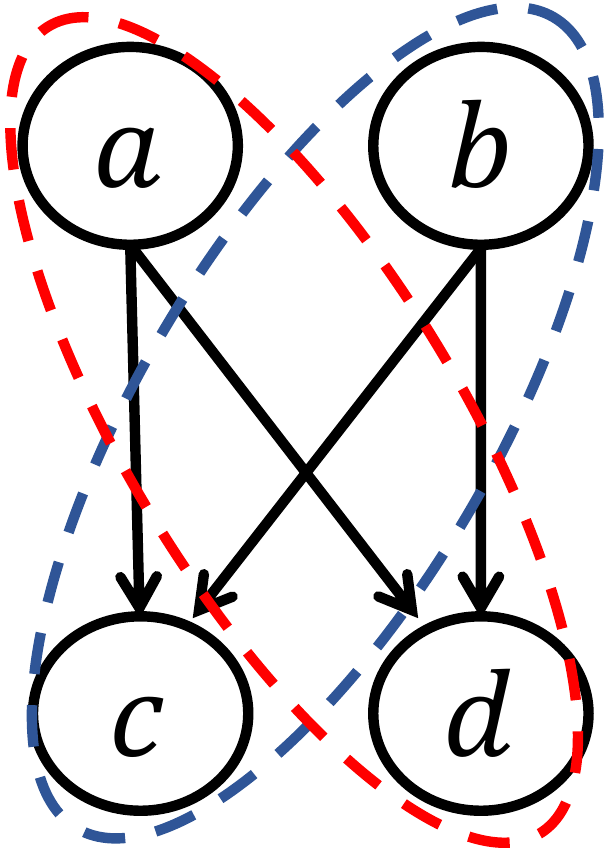}
 \caption{A cyclic partition}\label{fig:conv:3}\end{subfigure}
 \begin{subfigure}[t]{.25\linewidth} \includegraphics[width=\textwidth]{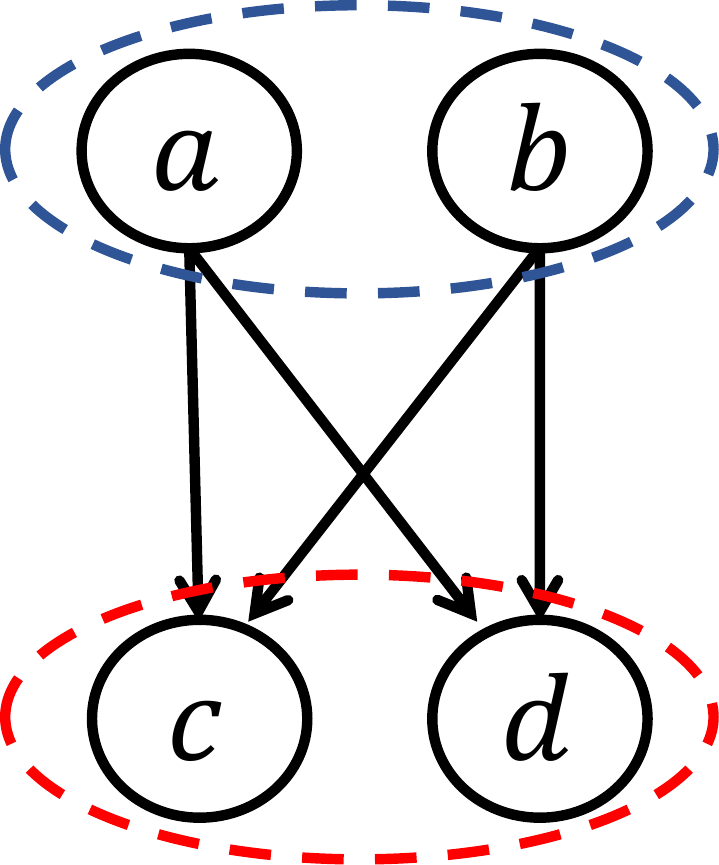}
 \caption{An acyclic partition}\label{fig:conv:4}\end{subfigure}
\caption{A toy graph (left), two cyclic and convex partitions (middle two),
      and an acyclic and convex partition (right).}
\label{fig:convex}
\end{figure}

\Cref{fig:convex} presents examples and difference of cyclic and acyclic partitions
for the toy graph shown in \cref{fig:conv:1}, there are three possible balanced partitions
shown in \cref{fig:conv:2}, \cref{fig:conv:3}, and~\cref{fig:conv:4}. The quotient graphs for
both~\cref{fig:conv:2,fig:conv:3} contains two nodes: red and blue, and two edges with cost 1.
In~\cref{fig:conv:2}, the edge from $a$ to $d$ creates an edge from blue to red, and the edge between
$b$ and $c$ creates and edge from red to blue, creating a cycle in the quotient graph.

\section{Related work}
\label{sec.related}

The recent work of Henzinger et al.~\cite{henzinger2020ilp} presents a formulation for the
undirected balanced graph partitioning problem. This model includes the essential definition and
constraints, and can be used as a basis for our problem variant.
The recent notable efforts on mathematical formulation of acyclic partitioning problem are due to
Nossack et al.~\cite{nope:14,nossack2014mathematical} and Albareda-Sambola et al.~\cite{albareda2019reformulated}.

Nossack et al.~\cite{nope:14} introduce a model with a high memory complexity. Their follow-up work~
\cite{nossack2014mathematical} improves this model, and introduces another novel model from a
different perspective, augmented set partitioning formulation.
Both formulations by Nossack et al.~\cite{nossack2014mathematical,nope:14}
rely on the Miller-Tucker-Zemlin (TMZ) subtour elimination
constraints from the Traveling Salesperson Problem~\cite{miller1960integer}
for the acyclicity constraint.
As the problem is dividing the set of vertices into subsets, the resulting assignment may present
equivalent (e.g., symmetric) optimal solutions from the possible solutions space.
Thus, they introduce additional constraints to reduce the symmetry.
We further discuss one of the models they presented in~\cref{sub.nossackmodel1}.

Albareda-Sambola et al.~\cite{albareda2019reformulated} reformulates the model presented by Nossack
et al.~\cite{nossack2014mathematical}, and introduces a rather comprehensive preprocessing of
ancestor and descendant sets of each node to forbid beforehand some node pairs that cannot lead to a valid
partitioning and introduces additional valid inequalities as constraints in order to limit the
search space and speed up the execution time of ILP solvers for their formulation.

Their formulation starts with the same base constraints and
then moves towards a topological-order-based formulation approach,
which is a huge step towards a simpler formulation.
On the other hand, the inequalities presented as constraints include many complex components and the
preprocessing-related variables, which makes the final set of constraints hard to
understand (e.g., pairwise connectivity of all nodes, $O(|V| \cdot |E|)$ pre-computed values, many constraints, etc.). We discuss their formulation further in~\cref{sub.albaredamodel}.

Before delving into acyclic partitioning formulations,
we first briefly introduce the mathematical formulation of the undirected balanced graph
partitioning from Henzinger et al.~\cite{henzinger2020ilp} in~\cref{sub:undirectedmodel},
since it introduces the common components of all formulations.
Then, in~\cref{sub.nossackmodel1}, we present the improved formulation model by Nossack et
al.~\cite{nossack2014mathematical}, and in~\cref{sub.albaredamodel}, we briefly describe the reformulation by
Albareda-Sambola et al.~\cite{albareda2019reformulated}.

\subsection{Formulation for Undirected Graph Partitioning} 
\label{sub:undirectedmodel}

The first step is to introduce a variable for the computation of the objective function, i.e., edge cut.
$z_{ij}$ is defined for each edge as a binary variable which is set to one if the edge $(i,j)$ is cut,
and zero otherwise. Then, the objective funtion is defined as the minimization of the sum of the
cost of cut edges (eq. \ref{eq:und:obj}).
An equivalent definition sets $z_{ij}$ as zero for cut edges and one for internal (uncut) edges and
defines the objective function as maximizing the sum of the cost of internal edges (alternatively
phrased as $z_{ij}$ is set to 1 if $i$ and $j$ belong to the same part, and zero otherwise),
which is the preferred presentation by both Nossack et al.~\cite{nossack2014mathematical}
and Albareda-Sambola et al.~\cite{albareda2019reformulated}.

The constraints (\ref{eq:und:constr_onepart}) enforce
that each node is assigned to exactly one part, and (\ref{eq:und:constr_partbal}) enforce the total
weight balance constraint for each part. The constraints (\ref{eq:und:constr_samepart1})
make sure that the $z_{ij}$ variables are set to 1 if the vertices $i$ and $j$ are not assigned to the same
part. $B$ denotes the balance weight limit including the imbalance ratio factored in as defined in~\cref{sec:preliminaries}.
The $w_i$ denotes the weight of vertex $i$, $c_{ij}$ denotes the cost of edge $\{i,j\}$ and defined
only over the set of existing edges.
The decision variable $x_{is}$ is a binary variable set to 1 if the vertex $i$ is assigned to the
part $s$.

Then, the formulation is as follows. \\
Objective: Minimize the edge cut.
\begin{equation}
\label{eq:und:obj}
min \sum_{(i,j) \in E} z_{ij} \cdot c_{ij}
\end{equation}
Subject to:
\begin{align}
\shortintertext{Constraint that all nodes belong to exactly one part:}
&\sum_{s = 0}^{k-1} x_{is} = 1 &\forall i \in V, \quad 0 \leq i < N = |V| \label{eq:und:constr_onepart}\\
\shortintertext{Constraint for part weight balance:}
&\sum_{i \in V} w_i \cdot x_{is} \leq B &\forall 0 \leq s < k \label{eq:und:constr_partbal}\\
\shortintertext{Constraint for marking (i.e., setting the $z_{ij}$ variable to one) the cut edges if
they are in the different parts:}
&z_{ij} \geq |x_{is} - x_{js}| &\forall \{i, j\} \in E, 0 \leq s < k \label{eq:und:constr_samepart1}\\
\shortintertext{Domains of decision variables:}
& x_{is} \in \{0,1\} &\forall i \in V, 0 \leq s < k \label{eq:und:dom_x}\\
& z_{ij} \in \{0,1\} &\forall \{i,j\} \in E, 0 \leq s < k \label{eq:und:dom_z}
\end{align}

In all following formulations with the same objective (i.e., minimizing the edge cut/maximizing the weight of
internal edges), the $z_{ij}$ variables can be relaxed to continuous variables.


\subsection{Nossack et al.'s Model}
\label{sub.nossackmodel1}

Nossack et al.~\cite{nope:14}'s initial formulation deviates from the formulation
defined in \cref{sub:undirectedmodel} by defining the z variable
three dimensional in which the first two dimensions are for the pairs of nodes, and the third is for a specific part number.
Their next work improves this initial formulation and uses a two dimensional z variable. Hence, we
ignore their earlier formulation and focus on the improved formulation presented in~\cite{nossack2014mathematical}.

The main components of the formulation does not deviate drastically from the undirected partitioning to acyclic DAG partitioning.
Only, the acyclic DAG partitioning includes additional constraints to address the acyclicity. And
several constraints to further decrease the search space and improve runtime performance in ILP solvers.
The mathematical formulation of Nossack et al.'s approach assumes the number of parts can be as
high as the number of nodes. To unify and simplify the presentation, we present the limit of number of parts
as $k$, as opposed to $N = |V|$.
As noted, the definition of $z_{ij}$ is the reverse of Henzinger et al.~\cite{henzinger2020ilp}'s,
i.e., the cut edges are marked as zero and internal (uncut) edges are marked as one.

Their formulation is as follows:\\
Objective: Maximize the total cost of internal edges, i.e., edges whose end points $i$ and $j$ belong to the same part $s$.
\begin{equation}
\label{eq:obj1}
max \sum_{(i,j) \in E} z_{ij} \cdot c_{ij}
\end{equation}
Subject to:
\begin{align}
\shortintertext{Constraint for all nodes belonging to exactly one part:}
&\sum_{s = 0}^{k-1} x_{is} = 1 &\forall i \in V \\
\shortintertext{Constraint for part weight balance:}
&\sum_{i \in V} w_i \cdot x_{is} \leq B &\forall 0 \leq s < k\\
\shortintertext{Constraint for marking the internal (uncut) edges if they are in the same part:}
&z_{ij} + x_{is} - x_{js} \leq 1 &\forall i < j, \{i,j\} \subset V, 0 \leq s < k \label{eq:nos:constr_samepart1}
\end{align}
Constraints for applying triangular inequality for all nodes $j$ that lie in any path between $i$
and $h$. If $i$ and $h$ are in the same part, $j$ must be in the same part as well. If $i$ and $j$
are not in the same part (i.e., $z_{ij}$ is cut), $i$ and $h$ cannot be in the same part (i.e.,
$z_{ih}$ must also be cut), and if $j$ and $h$ are not in the
same part, $i$ and $h$ cannot be in the same part:
\begin{align}
& \begin{rcases}
&z_{ij} + z_{jh} - z_{ih} \leq 1 \\
&z_{ij} - z_{jh} + z_{ih} \leq 1 \\
&-z_{ij} + z_{jh} + z_{ih} \leq 1 \\
&z_{ih} \leq z_{ij} \\
&2z_{ih} \leq z_{ij} + z_{jh}
\end{rcases} & &\forall i < j  < h, \{ i,j,h\} \subset V, i \leadsto j, j \leadsto h \label{eq:nos:constr_tri1}\\
\end{align}
Constraints for variables in $y$ matrix, i.e., induced edges represented as an adjacency matrix of
parts, i.e., every cell of $y$ matrix, $y_{st}$ has value 1 if there exists an edge between any node
in $V_s$ to any node in $V_t$:
\begin{align}
&x_{is} + x_{jt} - 1 \leq y_{st} &\forall (i,j) \in E, i \in V_s, j \in V_t, 0 \leq s \neq t < k
\label{eq:nos:constr_induced_edge}\\
\shortintertext{Constraints for $\pi$ values from TMZ subtour (cycle) elimination constraints (See ~\cite{miller1960integer}):}
&|V| \cdot (y_{st} - 1) \leq \pi_t - \pi_s - 1 &\forall 0 \leq s \neq t < k \label{eq:nos:constr_pi}\\
\shortintertext{Constraint to decrease symmetry by assigning the part indices sorted by total
vertex weight within the part:}
& \sum_{i \in V} x_{is} \leq \sum_{i \in V} x_{i,s-1} &\forall 1 \leq s < k
\label{eq:nos:constr_symdec}\\
\shortintertext{Domains of decision variables:}
& x_{is} \in \{0,1\} &\forall i \in V, 0 \leq s < k \label{eq:nos:dom_x}\\
& z_{ij} \in \{0,1\} &\forall i < j, \{i, j\} \subset V \label{eq:nos:dom_z}\\
& y_{st} \in \{0,1\} &\forall 0 \leq s \neq t < k \label{eq:nos:dom_y}\\
& \pi_{s} \in \mathbb{Z} &\forall 0 \leq s < k \label{eq:nos:dom_pi}
\end{align}

Although an essential part of the formulation, the constraints that address the acyclicity are
adapted from Miller-Tucker-Zemlin subtour elimination formulation for Traveling Salesperson
Problem~\cite{miller1960integer}.

The essence of the approach is to define an integer $\pi$ variable for each part within the range
[0,k) where each part is to be assigned a unique value. For a valid acyclic partitioning there
exists a one-to-one mapping between the integers in the range [0,k) and the $\pi$ values. The $y$
matrix is essentially the adjacency matrix for the induced, quotient graph.
The $\pi$ and $y$ variables together with the constraints in (\ref{eq:nos:constr_tri1}),
the formulation enforces the parts to have unique topological order indices and all nodes that lie
in between two nodes from the same part in any path to be assigned in the same part.

Since the part assignments can lead to many symmetrical solutions (e.g., just the possible reorderings of
part indices lead to $k!$ identical partitions), the constraint (\ref{eq:nos:constr_symdec}) is added to enforce
part indices to start from the largest part to smallest in total vertex weight size. Although this
does not prevent all symmetrical solutions, it reduces the optimal solution count significantly.

\subsection{Albareda-Sambola et al.'s Model}
\label{sub.albaredamodel}

Albareda-Sambola et al.~\cite{albareda2019reformulated}'s main idea is to improve upon the
previous
formulations by filtering out some of the impossible cases with the help of a pre-processing phase and
introducing additional valid inequalities. They experiment with many valid inequalities as
the set of constraint combinations.
The new formulations they present have closer ties to the topological orderability of DAGs.
The formulation enforces that the part assignment follows a topological order, i.e., there
exists an edge $(i,j) \in E, i \in V_s, j \in V_t$ if and only if $s < t$.
That is, the indices of the parts should be sorted in a topological order.
This is a huge step towards a simpler formulation of the model. However, combined with the
complexity of the variables introduced in pre-processing and constraints defined as a sum function
on the part assignment vectors for each pair of connected nodes
makes the formulation more complex and hard to follow.

Their pre-processing step defines the new variables $\alpha_{ij} = 1$ if $i \leadsto j$ and
$A_{ij}$ for each $(i,j)$ pair where $i < j, i \leadsto j$ as
$$A_{ij} = w_i + w_j + N_i^j.$$
Last,
$A'_{ijl}$ is defined as the weight sum of all distinct nodes on all paths $i \leadsto j$, $j
\leadsto l$, and $i \leadsto l$, i.e., $$A'_{ijl} = w_i + w_j + w_l + \sum \limits_{h \in N_i^j \cup N_j^l \cup N_i^l
\backslash \{j\}} w_h.$$

Here, the $\alpha_{ij}$ variable is equal to one if $j$ is a descendant of $i$,
i.e., there exists a path from $i$ to $j$. And, the $A_{ij}$ variables store the weight sum of all nodes
that lie on any path between the nodes $i$ and $j$, including $i$ and $j$.
The intuition behind the $A_{ij}$ variables is that for any acyclic partitioning
in which nodes $i$ and $j$ are assigned to the same part, all nodes that are on any path between the
two nodes must be in the same part as well. Thus, if $A_{ij} > B$, there is no feasible solution
that assigns $i$ and $j$ to the same part.
Computation of total node weight $A_{ij}$ for each pair of ancestor-descendant nodes helps
eliminate part assignments where the total part weight constraint is violated.
The downside, on the other hand, is the computation of all distinct vertices on any path between all $(i,j)$
connected pairs of nodes and the introduction of $\mathcal{O}(k \cdot |V|^2)$ constraints as in
(\ref{eq:alba:weight_res_topo1}) and (\ref{eq:alba:weight_res_topo2}) can be computationally expensive and
challenging. For small graphs, this can be a trivial
computation but as the scale and density of graphs increase, the amount of data to store and compute
increases drastically, leading to the need for more complex algorithms that can do those
computations efficiently.

The formulation of Albareda-Sambola et al. is as follows.\\
Objective: Maximize the sum of the cost of internal edges.
\begin{equation}
\label{eq:obj2}
max \sum_{(i,j) \in E}^{} c_{ij} \cdot z_{ij}
\end{equation}
Subject to:
\begin{align}
\shortintertext{Constraint for all nodes belonging to exactly one part:}
&\sum_{s = 0}^{k-1} x_{is} = 1 &\forall i \in V \label{eq:alba:constr_onepart}\\
\shortintertext{Constraint for part weight balance:}
&\sum_{i \in V} w_i \cdot x_{is} \leq B &\forall 0 \leq s < k \label{eq:alba:constr_partbal}
\end{align}
The following constraints are to preserve the acyclicity of partitioning and the topological order
of part indices.
These constraints for part assignment prevent creation of edges $(V_s,V_t)$ where $s>t$ in the quotient graph adjacency matrix:
\begin{align}
&  \sum \limits_{t \geq s} x_{it} + \sum \limits_{t < s} x_{jt} \leq 1 & \alpha_{ij} = 1,  A_{ij} \leq B, \forall 0 \leq s < k  \label{eq:alba:weight_res_topo1}\\
&  \sum \limits_{t \geq s} x_{it} + \sum \limits_{t \leq s} x_{jt} \leq 1 & \alpha_{ij} = 1,  A_{ij} > B, \forall 0 \leq s < k  \label{eq:alba:weight_res_topo2}\\
&  z_{ij} + \sum \limits_{t < s} x_{it} + \sum \limits_{t \geq s} x_{jt} \leq 2 &
\begin{array}{r}
\alpha_{ij} = 1, A_{ij} > B,\\
\forall (i,j) \in A, \forall 0 \leq s < k
\end{array}\label{eq:alba:weight_res_topo3}\\
\shortintertext{Domains of decision variables:}
& z_{ij} \in \{0,1\}  &\forall (i,j) \in E, \{i,j\} \subset V \label{eq:alba:dom_z}\\
& x_{is} \in \{0,1\} &\forall i \in V, 0 \leq s < k \label{eq:alba:dom_x}
\end{align}

The acyclicity constraints in (\ref{eq:alba:weight_res_topo1} - \ref{eq:alba:weight_res_topo3})
above restrict the part assignments of $i$ and $j$ to $x$ and $y$, $x \leq y$ if $A_{ij} \leq B$ and
$x< y$ if $A_{ij} > B$ respectively. The constraint (\ref{eq:alba:weight_res_topo3}) ensures the
correct assignment of $z_{ij}$ variables when $i$ and $j$ are assigned different parts.
Note that although the set of decision variables is smaller, the preprocessed $A_{ij}$ values may be
as many as $\mathcal{O}(|V|^2)$.

The above formulation is complete by itself, however,
to improve the formulation, the authors add further filtering constraints using the computed $A_
{ij}$ and $A'_{ijl}$ values from pre-processing and add the following constraint to make their
final best performing formulation:
\begin{align}
& z_{ij} = 0 & \forall (i,j) \in A, A_{ij} > B.
\end{align}

In addition, modify the triangular inequality constraints to utilize the $A_{ij}$ and B values as follows:

\begin{align}
& \begin{rcases}
        & z_{il} \geq z_{ij}  + z_{jl} - 1 \\
        & z_{ij} \geq z_{il}  + z_{jl} - 1 \\
        & z_{jl} \geq z_{ij}  + z_{il} - 1 \\
    \end{rcases} & \forall (i,j) \in A, A_{ij} > B
\end{align}
\begin{align}
    & z_{il} \leq z_{ij} & \forall i<j<l \in V, \quad A_{ij}, A_{il} \leq B, j \in N_l^i \\
    & z_{il} \leq z_{jl} & \forall i<j<l \in V, \quad A_{ij}, A_{jl} \leq B, j \in N_l^i
\end{align}
\begin{align}
    & z_{ij} + z_{jl} + z_{il} \leq 1 & \forall i<j<l \in V, \quad A_{ij}, A_{il}, A_{jl} \leq B, A'_{ijl} > B \\
    & z_{ij} + z_{il} \leq 1 & \forall i<j<l \in V, \quad A_{ij}, A_{il} \leq B, A_{jl} > B \\
    & z_{il} + z_{jl} \leq 1 & \forall i<j<l \in V, \quad A_{il}, A_{jl} \leq B, A_{ij} > B \\
    & z_{ij} + z_{jl} \leq 1 & \forall i<j<l \in V, \quad A_{ij}, A_{jl} \leq B, A_{il} > B
\end{align}
Finally, as the final extension and simplification to their formulation, they replace the constraints
(\ref{eq:alba:weight_res_topo1}-\ref{eq:alba:weight_res_topo3}) with the following:
\begin{align}
    & w_i + \sum \limits_{i-1}^{j=1} w_jz_{ji} + \sum \limits_{j=i+1}^{|V|-1} w_jz_{ij} \leq B &
    \forall 0 < i < N\label{eq:alba:monster_weight_const} \\
    & z_{ij} + z_{il} + \sum \limits_{t \geq s} (x_{jt} + x_{lt}) + \sum \limits_{t<s} x_{it} \leq 3 &
    \begin{array}{r}
    \forall i < j < l \in V, \alpha_{jl} = 0,\\
    A_{ij},A_{il} \leq B, A'_{ijl} > B, \forall 0 \leq s < k
    \end{array}
    \label{eq:alba:monster_acyc1}\\
    & \sum \limits_{t \geq s} x_{it} + \sum \limits_{t \leq s} x_{jt} \leq 1 + z_{ij} &
    \begin{array}{r}
    \forall 0 < i < j \in V,\\
    A_{ij} \leq B, A'_{ijl} > B, \forall 0 \leq s < k
    \end{array}\label{eq:alba:monster_acyc2}
\end{align}

\section{A simple and elegant formulation}
\label{sec.algo}

Our formulation builds on top of the simple idea of topological orderability of DAGs.
It is known that for any given DAG there exists at least one topological ordering of the nodes in
which all edges are from lower order nodes towards higher order nodes,
and, a graph is topologically orderable if and only if it is a directed acyclic graph.
This has been used in many different
domains that makes use of DAGs such as dynamic topological order maintenance algorithms for
streaming and evolving DAGs~\cite{pearce2007dynamic}, and indeed, Albareda-Sambola et al.~
\cite{albareda2019reformulated} makes use of this property in their formulation as well.

In our formulation, we return to the roots of the problem, and build our solution with the minimal
and simple constraint inequalities: Given a DAG $G$, we want to partition the vertex set into
disjoint subsets where the quotient graph is also a directed acyclic graph. Therefore, the resulting quotient graph should also have at least one topological ordering.
Then, we can ignore the many symmetrical solutions by focusing on one specific solution where the part indices
are assigned in one arbitrary valid topological ordering for the quotient graph, and enforce this information at the
\emph{adjacency matrix} of the quotient graph.
Mathematically, the formulation becomes enforcing the adjacency matrix to be an upper triangular matrix.

Our proposed formulation is as follows.\\
Objective: Minimize the edge cut where $z_{ij}$ is one for a cut edge and zero otherwise.
\begin{equation}
\label{eq:obj3}
min \sum_{(i,j) \in E}^{} c_{ij} \cdot z_{ij}
\end{equation}
Subject to:
\begin{align}
\shortintertext{Constraint for all nodes belonging to exactly one part:}
&\sum_{s = 0}^{k-1} x_{is} = 1 &\forall i \in V \label{eq:my:constr_onepart}\\
\shortintertext{Constraint for part weight balance:}
&\sum_{i \in V} w_i \cdot x_{is} \leq B &\forall 0 \leq s < k \label{eq:my:constr_partbal}\\
\shortintertext{Constraint for marking the cut edges as 1 if they are in different parts:}
&x_{js} - x_{is} \leq z_{ij} &\forall (i,j) \in E, 0 \leq s < k \label{eq:my:constr_samepart} \\
\shortintertext{Constraints for the decision variables in $y$ matrix, i.e., induced edges represented as an adjacency matrix of parts:}
&x_{is} + x_{jt} -1 \leq y_{st}
& \begin{array}{r}
\forall (i,j) \in E, i \in V_s, j \in V_t,\\
0 \leq s \neq t < k
\end{array}
\label{eq:my:constr_induced_edge}\\
\shortintertext{Constraint for topologically ordering the part indices and decrease symmetry by
restricting the strictly lower triangle of the y matrix:}
& y_{st} = 0 &\forall 0 \leq t < s < k
\label{eq:my:constr_triangular}\\
\shortintertext{Domains of decision variables:}
& x_{is} \in \{0,1\} &\forall i \in V, 0 \leq s < k \label{eq:my:dom_x}\\
& z_{ij} \in \{0,1\} &\forall (i,j) \in E, \{i,j\} \subset V \label{eq:my:dom_z}\\
& y_{st} \in \{0,1\} &\forall 0 \leq s,t < k \label{eq:my:dom_y}
\end{align}

Here, the value assignments of $x_{is}$ and $z_{ij}$ variables shape the nonzero values of $y$ matrix.
The acyclicity constraint is enforced by the restriction on the strictly lower triangle of the $y$
matrix. This restriction indirectly prevents any assigment of parts where $(i,j) \in E, i \in
V_s, j \in V_t$ and $t < s$ since $y_{st} \geq 1$ contradicts with $y_{st} = 0$ in this case.
It is important to note that the $y$ variable in this formulation need not store the exact adjacency
matrix since there is no tight-bound constraint nor an objective function defined on it. Formally,
we can define the $Adj$, adjacency matrix for the part graph where:
\begin{align}
Adj_{st} = & \begin{cases}
        1 & \exists (i,j) \in E, i \in V_s, j \in V_t \\
        0 & otherwise
    \end{cases}
\end{align}
Then, $y_{st} \geq Adj_{st}$.
Thus, there is no restriction to set the
values to zero for the upper triangle of $y$ when there is no edge to indicate so in the quotient graph.
However, the constraints enforce that all nonzero cells of the actual adjacency matrix of the quotient
graph are correctly assigned nonzeros in the $y$ matrix as well. And, the critical section, i.e.,
the strictly lower triangle is always set to zero.

This formulation saves us from the need for additional variables around TMZ formulation (e.g.,
$\pi$) and constraints to eliminate symmetrical solutions as was used in Nossack et al.~
\cite{nossack2014mathematical}'s formulations as well as complex formulations with
significant pre-processing steps as was used in Albareda-Sambola et al.~
\cite{albareda2019reformulated}'s formulations.

\section{Real-World Impact} 
\label{sec:real_world_impact}

Since ILP formulations of the acyclic k-way partitioning problem are not feasible algorithms for large inputs (e.g., number of nodes $>$ 1000),
we focus on the scenarios where this deficiency can be mitigated.
Here, we give two examples of where this new formulation is useful. First, we design heuristic
algorithms that utilize ILP formulation within the popular multilevel acyclic partitioning
paradigm. Then, we present a real-world example use case, i.e., partitioning for hierarchical
state-vector based
quantum circuit simulations, where the ILP is used as the lower bound/optimal result as the baseline
for heuristic algorithms.

\subsection{Application within Multilevel Partitioning Paradigm} 
\label{sub:application_within_multilevel_partitioning_paradigm}

Multilevel partitioning is first introduced in 1990s~\cite{bui1993heuristic}, and is the de facto approach
for many graph partitioning problems~\cite{kaku:98:metis,Catalyurek13-UMPa,patohmanual,pell:08:scotch,Herrmann19-SISC,mops:17b}.
In high level, a multilevel partitioning algorithm consists of 3 phases: coarsening, initial
partitioning, and uncoarsening/refinement. Coarsening is the application of a series of contractions
of the nodes of an input graph to create smaller but similar instances of the input.
The goal is to reduce the size of the problem to a more manageable size
while preserving the main features of the input.
The initial partitioning phase is where the coarsest graph is partitioned into the desired
number of parts. The initial partitioning can afford expensive algorithms
that would not be feasible for the original input
since the coarsening phase ideally reduces the problem size significantly.
The uncoarsening phase is essentially the reverse of the coarsening phase:
The small, coarse graph is uncontracted back to the original layer by layer. And, at each step of
the uncoarsening, a refinement algorithm is applied in order to improve the objective function.

There are multiple ways to apply the presented simplified perspective to the acyclic
partitioning problem to the multilevel paradigm. We briefly mention three opportunities: We can
\begin{enumerate}
\item design coarsening and refinement algorithms that
maintain acyclicity and the node indexing which conforms to the topological ordering. This idea is
closely related with the application of dynamic topological order maintenance
algorithms~\cite{pearce2007dynamic} to the partitioning result, however,
it is not implemented in the recent acyclic graph partitioning algorithms~\cite{Herrmann19-SISC,mops:17b}.
Maintaining a topological order of partitions/contracted nodes, or, maintaining an
upper-triangular matrix of adjacency relation, during coarsening and refinement
allows us to reduce the calls to cycle detection procedures.
We would need to run it only for the cases where
the change in the graph creates an edge from a higher indexed node (or part) to a lower indexed node (or part).
Thus, effectively reduces the potential necessary cycle detection calls by half
since creation of any edge from a lower indexed entity to a higher indexed entity
as well as removal of any edge cannot create a cycle.

\item define an ILP-based initial partitioning
(either as an optimal partitioning or a time-restricted heuristic approach),
which may be feasible only because coarsening can bring the size of the graph to a manageable size.
Normally, ILP solutions are not feasible for larger instances because ILP defines exponentially more
variables and constraints which make it much harder to store and process.
Since typically, coarsening phase is used to produce small representatives of input (e.g., less than 500 nodes),
it may bring the execution time within a tolerable range.

\item use ILP formulation as a refinement algorithm,
using the result of initial partitioning as a hot start.
Many recent ILP solvers allow starting with a user-defined valid or partial solution.
And, users have the option to either search for the optimal solution or search for an improvement given a time limit.
As with any other refinement algorithm, one can project back the partitioning of coarser graph to
the current,
finer graph and use this partition projection as initial solution for the ILP solver.
Although the ILP solvers do not make any promises about the use of or improvement upon a given initial solution,
the result would be at least as good as the given initial solution.
\end{enumerate}
Exploring the best ways to design coarsening and refinement phases with ILP solvers for acyclic partitioning is an interesting, ongoing research effort.
Algorithms toward similar goals are developed for the general undirected graph partitioning problem in~\cite{henzinger2020ilp}.


\subsection{Application in Quantum Circuit Simulation Problem} 
\label{sub:application_in_quantum_circuit_simulation_problem}

Using ILP for partitioning is generally not efficient since the problem sizes can get quite large
and ILP solvers are not favorable in terms of the runtime in this case.
Quantum circuit simulation is a recent problem area
where a quantum computation is represented as
a directed acyclic graph and simulated using classical computers.
Currently, the largest quantum computers can compute circuits with up to 127 qubits (quantum
bits), however, many simulators can not handle even the half of this number~\cite{Fang22-ARXIV}.
As the number of qubits are still low, the simulated circuits are also quite small compared to
inputs of other use cases (e.g., simulations and scheduling for classical computation~
\cite{Ozkaya19-IPDPS,Ozkaya19-PPAM}). This makes ILP solver based
partitioning a feasible approach for the partitioning of quantum circuit simulation algorithms.

We develop an ILP formulation for the specific partitioning problem defined in HiSVSIM~\cite{Fang22-ARXIV}.
Here, the problem is partitioning a quantum circuit DAG acyclically into
minimum possible number of parts,
where the resulting parts contain no more than a given number of unique qubits.
The added constraints are for counting and limiting the unique qubits per part given a
maximum allowed number of qubits limit ($L_m$).
Finally, the problem objective is not only the minimization of edge cut, but also the number of parts.

The input datasets for HiSVSIM~\cite{Fang22-ARXIV} consists of 13 quantum circuits (9 unique quantum circuits)
where the number of qubits are between 30 and 37.
Out of 13 circuits,
10 of them contain less than 500 quantum gates,
and 6 of them contain less than 200 gates.
Thus, the problem sizes are small enough to try ILP solver based approaches.

The input graphs for this problem has the following structure.
All quantum gates are represented as nodes
and all qubits (operands of quantum computation gates) are represented as edges.
The graph contains \emph{entry} and \emph{exit} gates for each qubit.
Each qubit is an in-edge to and out-edge from a single node at any time.
And, qubits can be traced as a line subgraph from their respective entry nodes to respective exit nodes.
Thus, given a quantum circuit DAG, the involved qubits of each node is known.
And given a subset of nodes, it is trivial to identify the unique qubits involved in this subset.

We define $Q$ as a set of qubits, binary $NQ$ matrix to store node(quantum gate)-qubit dependence and
the values of $NQ$ are pre-computed in linear time with respect to the number of edges in the DAG.
The cell $NQ_{iq}$ stores 1 if the qubit $q$ is required for the computation of node $i$.
Then, we define a binary decision variable $PQ$ that stores the qubit dependence of parts, i.e.,
the cell $PQ_{sq}$ stores 1 if the qubit $q$ is required for part $s$.
Then, the ILP formulation for this problem variant becomes:

Objective: Minimize the edge cut where $z_{ij}$ is one for a cut edge and zero otherwise.
\begin{equation}
\label{eq:obj4}
min \sum_{(i,j) \in E}^{} c_{ij} \cdot z_{ij}
\end{equation}
Subject to:
\begin{align}
\shortintertext{Constraint for all nodes belonging to exactly one part:}
&\sum_{s = 0}^{k-1} x_{is} = 1 &\forall i \in V \label{eq:my:q:constr_onepart}\\
\shortintertext{Constraint for part weight balance:}
&\sum_{i \in V} w_i \cdot x_{is} \leq B &\forall 0 \leq s < k \label{eq:my:q:constr_partbal}\\
\shortintertext{Constraint for marking the cut edges as 1 if they are in different parts:}
&x_{js} - x_{is} \leq z_{ij} &\forall (i,j) \in E, 0 \leq s < k \label{eq:my:q:constr_samepart} \\
\shortintertext{Constraints for the decision variables in $y$ matrix, i.e., induced edges represented as an adjacency matrix of parts:}
& x_{is} + x_{jt} -1 \leq y_{st}
& \begin{array}{r}
\forall (i,j) \in E, i \in V_s, j \in V_t,\\
0 \leq s \neq t < k
\end{array}
\label{eq:my:q:constr_induced_edge}\\
\shortintertext{Constraint for topologically ordering the part indices and decrease symmetry by
restricting the strictly lower triangle of the y matrix:}
& y_{st} = 0 &\forall 0 \leq t < s < k
\label{eq:my:q:constr_triangular}\\
\shortintertext{Constraints for limiting the number of qubits per part  by $L_m$:}
& PQ_{iq} \geq X_{is} \cdot NQ_{iq} &\forall i \in V, 0 \leq s < k, q \in Q \label{eq:my:q:quantum1}\\
&\sum_{q \in Q} PQ_{sq} \leq L_m &\forall 0 \leq s < k \label{eq:my:q:quantum2}\\
\shortintertext{Domains of decision variables:}
& x_{is} \in \{0,1\} &\forall i \in V, 0 \leq s < k \label{eq:my:q:dom_x}\\
& z_{ij} \in \{0,1\} &\forall (i,j) \in E, \{i,j\} \subset V \label{eq:my:q:dom_z}\\
& y_{st} \in \{0,1\} &\forall 0 \leq s,t < k \label{eq:my:q:dom_y} \\
& PQ_{sq} \in \{0,1\} &\forall 0 \leq s < k, q \in Q \label{eq:my:q:dom_quantum}
\end{align}

It is important to note that the objective in this problem is to minimize the number of parts, and
then the edge cut. There are two ways to achieve this. First, we can try, starting from a small k
value and increase one by one while trying to find a feasible solution and stop as soon as a
feasible solution is found (or binary search for the smallest k value with a feasible solution). Second, we can define an additional binary variable for each part that stores true if
a part contains at least one node. Then, multiply this variable with a sufficiently large number and
use it as an additive component in the objective function.
%
The proposed ILP formulation was successfully used to find the optimal partitioning solution for
the acyclic quantum circuit partitioning problem.


\section{Discussion and Conclusion}
\label{sec.conc}

Nossack et al.~\cite{nope:14} presents one of the few analyses of the
mathematical formulations for the acyclic partitioning problem.
Albareda-Sambola et al.~\cite{albareda2019reformulated} introduce
a pre-processing phase and many carefully deduced valid inequality constraints
to speed up the computation of an optimal solution when using linear solvers.
In this work, we present a formulation that can be used together with/in addition to the previous
formulations.
We would like to note that, although the main goal of this work is on
presenting a simple and elegant formulation,
our experiments on a set of DAGs using the latest version of
Gurobi Optimizer available today (v9.5.0)~\cite{gurobi} showed similar runtime
performance for our acyclicity constraints compared to Albareda et al.~\cite{albareda2019reformulated}'s formulations
for the ILP solver phase. And, that Gurobi Optimizer eliminates many
variables and constraints (rows and columns) as redundant during the its presolve phase for all
three formulations.
This indicates that the recent advances in the ILP solver software can make up for the need for
introducing additional valid inequalities as constraints to limit the search space.
This has been a significant tradeoff
decision for many for years as to whether and how many new decision variables and constraints to
include (which increases the model formulation complexity, but may reduce the solution search
space) for the perfect balance for the ILP runtime performance.

To conclude, we present a simple and elegant formulation for the acyclic DAG partitioning problem
that eliminates many additional variables, redundant constraints and symmetrical solutions, and,
finally we show two example real-world scenarios where an elegant ILP-based formulation can be
utilized.
Our formulation performs similarly compared to more complex formulations.
Finally, the simplicity of the formulation may help enable many others to easily define, model,
implement, and experiment with balanced acyclic DAG partitioning problem and formulations.


\end{document}